\documentclass[prl,aps,graphics,a4paper,10pt,twocolumn,nofootinbib]{revtex4-1}

\usepackage[margin=25mm]{geometry}
\usepackage{amsmath}
\usepackage{braket}
\usepackage{hyperref}
\usepackage{tikz}
\usepackage{pgfplots}
\usepackage{tikzscale}

\def\colorared{0.0}
\def\coloragreen{0.0}
\def\colorablue{0.0}

\def\colorbred{0.4}
\def\colorbgreen{0.4}
\def\colorbblue{1}

\def\colorcred{1}
\def\colorcgreen{0.6}
\def\colorcblue{0.6}

\newcommand{\be}{\begin{equation}}
\newcommand{\ee}{\end{equation}}

\newcommand{\prj}[1]{\ensuremath{| #1 \rangle \langle #1 |}}
\newcommand{\ovl}[2]{\ensuremath{\langle #1 | #2 \rangle}}
\newcommand{\lole}{\zeta}

\begin{document}

\title{Prominent interference peaks in the dephasing Anderson model}

\author{Yannic Rath}
\author{Florian Mintert}

\affiliation{Physics Department, Blackett Laboratory, Imperial College London, Prince Consort Road, SW7 2BW, United Kingdom}
\date{\today}

\begin{abstract}
The Anderson model with decoherence features a temporal evolution from localized eigenstates to a uniform spatial distribution bar any interference features.
We discuss the growth and decay of pronounced interference peaks on transient time-scales and develop an analytic understanding for the emergence of these peaks.
\end{abstract}

\maketitle

Interference is the distinctive feature that separates the physics of classical particles from wave-mechanics, in particular quantum mechanics.
Since interaction with an environment results in decoherence and limits the ability of any system to show interference effects, one can thus realize continuous transitions between quantum mechanical and classical regimes, parametrized by strength of the environment coupling or duration of exposure to the environment.

Typically the transition towards a classical regime implies that interference patterns are washed out such that maxima decrease and minima grow.
This is strictly true in the textbook example of the double-slit experiment, with an interference pattern resulting from two amplitudes only.
Patterns comprised of more amplitudes, however, can result in more complex transitions between quantum and classical, with structures in an interference pattern existing in intermediate regimes that exist neither in the perfectly quantum nor in the classical case~\cite{Ra1227}.

A particularly intricate interference effect is found in the Anderson model, where the destructive interference of many amplitudes
is reflected by exponentially localized eigenstates~\cite{PhysRev.109.1492,kramer1993localization}.
Decoherence will generally allow initially localized states to expand
as destructive interference is lifted~\cite{PhysRevLett.85.812,PhysRevA.82.042109}.
The details of such dynamics~\cite{PhysRevE.89.042129} and their stationary solutions~\cite{PhysRevLett.118.070402,0295-5075-119-5-56001} depend on the specific properties of the environment coupling, as encoded, for example, in a Lindblad operator.
Quite generically, decoherence rates are particularly high for states that can easily be distinguished by the environment,
whereas they tend to be low for states that can hardly be distinguished by the environment.
Decoherence, thus not necessarily just results in a pure attenuation of interference structures;
rather, populations can start to propagate as the coherence length shrinks, while interference can still exist on sufficiently small length scales.

We will show here that propagating populations can become trapped due to interference on small length scales, resulting in the emergence of sharp interference structures that decay only once the coherence length is sufficiently small for the system to approach its classical limit.

\section{Dephasing Anderson model}

The Anderson model~\cite{PhysRev.109.1492,kramer1993localization} for a single particle on a one-dimensional lattice with $N$ sites is defined in terms of the Hamiltonian
\begin{equation}
	H = \sum_{x} \epsilon_x \ket{x} \bra{x} + \mathcal{T} \, (\ket{x} \bra{x+1} + \ket{x+1} \bra{x}),\nonumber
\end{equation}
where $\ket{x}$ denotes the occupation of site $x$.
This Hamiltonian includes tunneling from sites $x$ to neighboring sites $x\pm 1$ with amplitude $\mathcal{T}$, and onsite energies $\epsilon_x$.
In contrast to translationally invariant systems, with delocalized eigenstates (Bloch waves), any small amount of disorder, {\it e.g.} in the onsite energies $\epsilon_x$, causes a transition to exponentially localized eigenstates~\cite{PhysRev.109.1492,kramer1993localization}.

Decoherence can be described in terms of a master equation with a Lindbladian $\mathcal{L}$.
We will consider Lindbladians satisfying
\begin{equation}
	\mathcal{L}(\ket{y}\bra{x}) = - \gamma \, f(x,y) \, \ket{y}\bra{x},
	\label{eq:lindblad}
\end{equation}
for all pairs of lattice sites $x$ and $y$,
with a fundamental decoherence rate $\gamma$, and a distance-dependent factor $f(x,y)$ with $f(x,x)=0$.
This Lindbladian induces only loss of phase coherence between different sites, but no dynamics of lattice populations.

The function $f(x,y)$ characterizes how well the environment coupling can resolve fine spatial structures in the system.
The extreme case $f(x,y)=1$ for $x\neq y$ describes a situation in which all spatial structures can be resolved equally well, but more realistic models would take into account a finite resolution with reduced decoherence rates for short spatial scales and a finite, maximum rate for asymptotically large separations \cite{PhysRevA.68.012105}.
Since the behavior for large separations will not be relevant for our purposes, we will use $f=f_q$ with $f_q=|x-y|^q$ and focus mostly on the case $q=1$.
In order to avoid ambiguities in the definition of the distance between two sites, we employ open and not the usual periodic boundary conditions.

\section{Dephasing dynamics}

\begin{figure*}
  \input{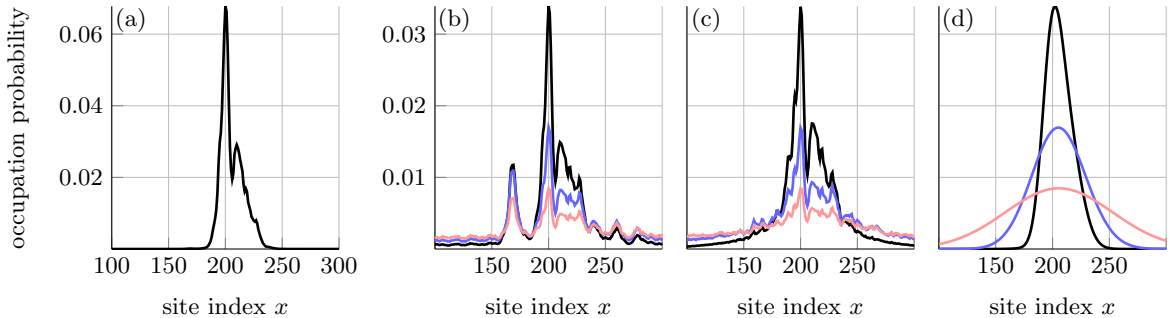}
	\caption{Time evolution for one particular realization of disorder.
	Figure (a) depicts the system ground state occupation, and (b) to (d) depict the time-evolved densities for three different dephasing rates, $\gamma = 10^{-9} \, \mathcal{T}$ in (b), $\gamma = 3 \cdot  10^{-3} \, \mathcal{T}$) in (c), and $\gamma = \mathcal{T}$ in (d).
	The three instances in time depicted in (b) to (d) are chosen such that the central peak has decayed to $1/2$, $1/4$ and $1/8$ of its original height (depicted in black (dark), blue (medium) and red (light)).
A clearly pronounced side-peak is growing and subsequently decaying in the case of weak dephasing (b), but intermediate and strong dephasing do not result in the emergence of such structures.}
	\label{fig:example_evolution}
\end{figure*}

In all the following analysis, we will discuss a lattice of $N=500$ sites, with onsite energies drawn from a uniform distribution in the interval $[-\mathcal{T}/10,\mathcal{T}/10]$,
but found the observed interference structures also for stronger disorder, including disorder substantially larger than $\mathcal{T}$.
Fig.~\ref{fig:example_evolution} depicts some explicit realizations of decoherence-induced dynamics with the ground state as initial state, for three different regimes of dephasing strength but the same realization of disorder giving rise to a particularly pronounced side-peak.
Since the time-scale on which interference structures decay depends non-linearly on the system parameters, the different points in time are not defined explicitly in terms of $\gamma$ and ${\cal T}$, but they are chosen such that the central peak has decayed to $1/2$, $1/4$ and $1/8$ of its original height in each of the sub-figures (b) to (d).
For strong dephasing (Fig.~\ref{fig:example_evolution}(d)) the initially localized peak widens and resembles qualitatively a Gaussian distribution that becomes broader as time evolves.
In an intermediate dephasing regime (Fig.~\ref{fig:example_evolution}(c)) the distribution of occupation broadens as a result of decoherence; fine structures give evidence of some coherent character of the dynamics but no large new structures arise.
This is fundamentally different in the slow dephasing regime, where
the coherent dynamics is much faster than the dephasing.
As shown in Fig.~\ref{fig:example_evolution}(b), an entirely new peak arises around the lattice site $x=169$.
At $t\mathcal{T}\simeq 8 \cdot 10^7$
(depicted in black (dark)) this peak is still nearly a factor $3$ smaller than the remains of the original peak, but whereas the height of the original peak decreases by a factor $4$, by the time $t\mathcal{T}\simeq 1.8\cdot 10^8$
(depicted in blue (medium)), the new peak hardly changes its height, and both peaks are almost of comparable height.

\begin{figure}
	\input{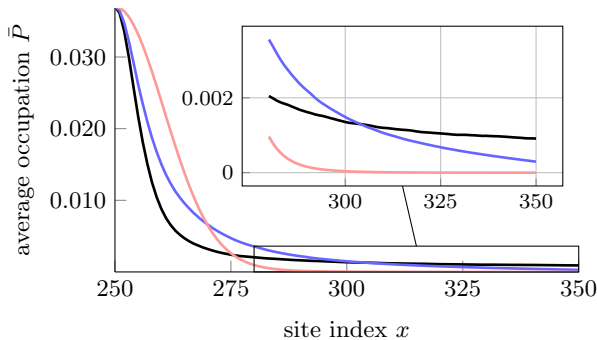}
	\caption{Ensemble average $\bar P(x)$ for the site occupation after an evolution time corresponding to a decay of the initial peak by a factor of $1/2$.
	Slow dephasing ($\gamma=10^{-9}\, {\cal T}$), intermediate dephasing  ($\gamma=3 \cdot 10^{-3}\, {\cal T}$) and fast dephasing  ($\gamma={\cal T}$) are indicated by black, blue and red (dark, medium and light) curves.
	Around the transition from peak to tail, there is an enhanced occupation for intermediate dephasing as compared to slow and fast dephasing.}
	\label{fig:ensemble_avg}
\end{figure}

Verifying that the qualitative observations made in Fig.~\ref{fig:example_evolution} are largely independent of the realization of disorder requires a proper statistical analysis.
Since different realizations of disorder result in ground states localized at different positions on the chain, we only consider realizations with ground states that have their center of mass in the interval $[249.5,250.5]$ around the center of the chain.
The resulting ensemble average $\bar P(x)$ for \(1000\) of such disorder realizations is depicted in Fig.~\ref{fig:ensemble_avg}, after a propagation time chosen for the three dephasing regimes such that the peak height of the ensemble average has reached half of its initial value.
As one can see, strong dephasing results in a rather broad peak but negligible tails, whereas weak dephasing yields more narrow peaks but pronounced tails.
Close to the center of the chain, {\it i.e.} close to the initial peak, and far out in the tails, the ensemble average for intermediate dephasing lies between the corresponding data for weak and strong dephasing;
between those regimes (in this case between approx. $x=270$ and $x=300$), however, the intermediate dephasing results in an average population $\bar P(x)$ that is larger than in the two extreme regimes.

The existence of a pronounced side-peak as visible in Fig.~\ref{fig:example_evolution} in the weak dephasing regime, is clearly consistent with the narrow peak and pronounced tails in the ensemble average shown in Fig.~\ref{fig:ensemble_avg},
but since the distance between side-peak and main peak depends on the realization of disorder, such side-peaks can not be unambiguously verified in the ensemble average.
We will therefore characterize peaks in interference patterns in terms of their topographic prominence~\cite{LLOBERA20011005}, which characterizes to what extent a peak stands out in front of the background.
In the present case of one-dimensional structures, the side prominence of a peak with respect to its left/right side can be defined as the difference between its height and the height of the lowest point between itself and the next higher peak to the left/right.
The overall prominence ${\cal P}_x$ for any peak at site $x$ is then given by the smaller value of the two side prominences.
We define the maximum side peak prominence $\Delta = {\max_{x, t_f}{\cal P}_x(t_f)-{\cal P}_x(0)}$ for a given system (characterized by realization of disorder and dephasing rate) as the maximal growth of the peak prominences over all times and all emerging peaks.

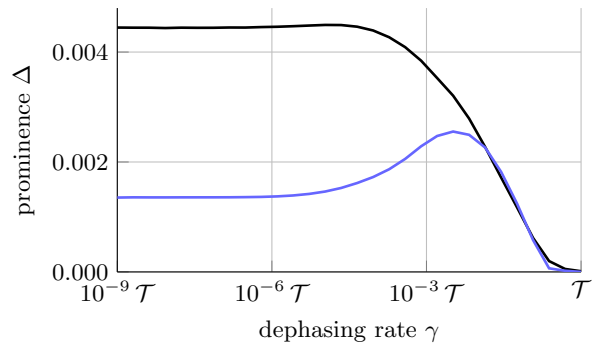
\begin{figure}
	%\begin{figure*}
\def\scale{0.5}
\begin{tikzpicture}[]
\begin{axis}[height = {\scale*101.6mm}, legend pos = {south west}, ylabel = {prominence $\Delta$}, xmin = {1.0e-9}, xmax = {1.0}, ymax = {0.0048}, xlabel = {dephasing rate $\gamma$}, unbounded coords=jump, scaled x ticks = false, xticklabel style={rotate = 0}, log basis x=10,
  xmajorgrids = true, xtick = {1.0e-9,1.0e-6,0.001,1.0}, xticklabels = {$10^{-9}\, \mathcal{T}$,$10^{-6}\, \mathcal{T}$,$10^{-3}\, \mathcal{T}$,$\mathcal{T}$}, xtick align = inside, axis lines* = left, scaled y ticks = false, yticklabel style={rotate = 0}, ymajorgrids = true, ytick = {0.0,0.002,0.004}, yticklabels = {0.000,0.002,0.004}, ytick align = inside, axis lines* = left,     xshift = 0.0mm,
    yshift = 0.0mm,
    axis background/.style={fill={rgb,1:red,1.00000000;green,1.00000000;blue,1.00000000}}
, xmode = {log}, ymin = {0}, width = {\columnwidth}]\addplot+ [color = {rgb,1:red,\colorared;green,\coloragreen;blue,\colorablue},
draw opacity=1.0,
line width=1,
solid,mark = none,
mark size = 2.0,
mark options = {
    color = {rgb,1:red,0.00000000;green,0.00000000;blue,0.00000000}, draw opacity = 1.0,
    fill = {rgb,1:red,0.17254902;green,0.49803922;blue,0.72156863}, fill opacity = 1.0,
    line width = 1,
    rotate = 0,
    solid
}]coordinates {
(1.0e-9, 0.004445569199696183)
(2.0433597178569395e-9, 0.004443366553611122)
(4.175318936560409e-9, 0.004442915710993111)
(8.531678524172815e-9, 0.0044369294903241095)
(1.743328822199987e-8, 0.004443446473171934)
(3.562247890262444e-8, 0.004441152836754918)
(7.278953843983146e-8, 0.004441697506606578)
(1.4873521072935117e-7, 0.0044465378512628375)
(3.0391953823131946e-7, 0.004445554618444294)
(6.210169418915616e-7, 0.00445436017587781)
(1.2689610031679234e-6, 0.0044600563059793786)
(2.592943797404667e-6, 0.004470260361395776)
(5.298316906283713e-6, 0.0044820584416855125)
(1.082636733874054e-5, 0.004493291346519254)
(2.21221629107045e-5, 0.004491113206441515)
(4.520353656360241e-5, 0.004461375752114691)
(9.236708571873866e-5, 0.004392192785255611)
(0.00018873918221350977, 0.004269358138204552)
(0.00038566204211634724, 0.004089392248657532)
(0.0007880462815669912, 0.0038425036075059324)
(0.0016102620275609393, 0.003527105847327039)
(0.0032903445623126675, 0.0032062241278123112)
(0.006723357536499335, 0.0027899964100914078)
(0.01373823795883263, 0.0022736543183680623)
(0.02807216203941177, 0.0017187836849829182)
(0.057361525104486784, 0.0011720317164901643)
(0.11721022975334802, 0.0006168036599524647)
(0.23950266199874853, 0.00019716292478053014)
(0.48939009184774934, 5.379704945190108e-5)
(1.0, 1.2504421042050935e-5)
};
\addplot+ [color = {rgb,1:red,\colorbred;green,\colorbgreen;blue,\colorbblue},
draw opacity=1.0,
line width=1,
solid,mark = none,
mark size = 2.0,
mark options = {
    color = {rgb,1:red,0.00000000;green,0.00000000;blue,0.00000000}, draw opacity = 1.0,
    fill = {rgb,1:red,0.98823529;green,0.55294118;blue,0.38431373}, fill opacity = 1.0,
    line width = 1,
    rotate = 0,
    solid
}]coordinates {
(1.0e-9, 0.0013527930034557358)
(2.0433597178569395e-9, 0.0013568500122055412)
(4.175318936560409e-9, 0.0013557920360006393)
(8.531678524172815e-9, 0.0013557340430095791)
(1.743328822199987e-8, 0.0013565901821479202)
(3.562247890262444e-8, 0.0013568274779245258)
(7.278953843983146e-8, 0.0013570277548162267)
(1.4873521072935117e-7, 0.0013578509010607376)
(3.0391953823131946e-7, 0.001360625852434896)
(6.210169418915616e-7, 0.0013647485115332529)
(1.2689610031679234e-6, 0.001374113185936585)
(2.592943797404667e-6, 0.001391029886342585)
(5.298316906283713e-6, 0.0014191027618013323)
(1.082636733874054e-5, 0.0014625793060520664)
(2.21221629107045e-5, 0.0015275886802235618)
(4.520353656360241e-5, 0.0016172831714735367)
(9.236708571873866e-5, 0.0017262660392560064)
(0.00018873918221350977, 0.0018678488634061069)
(0.00038566204211634724, 0.0020556684804847463)
(0.0007880462815669912, 0.002285907365451567)
(0.0016102620275609393, 0.0024729310711147263)
(0.0032903445623126675, 0.002553836398338899)
(0.006723357536499335, 0.0024942925431765614)
(0.01373823795883263, 0.00226318511296995)
(0.02807216203941177, 0.001819979826034978)
(0.057361525104486784, 0.0012402203446254135)
(0.11721022975334802, 0.0005788488967288733)
(0.23950266199874853, 6.55980685635203e-5)
(0.48939009184774934, 2.0716058598834484e-5)
(1.0, 1.753660804258062e-8)
};
\end{axis}

\end{tikzpicture}
%\end{figure*}
	\caption{Prominence as a function of dephasing rate \(\gamma\) for an ensemble of $5000$ realizations of disorder.
	The black/blue (dark/light) data correspond to sub-ensembles with the higher/lower prominence in the weak dephasing regime.
	There is a pronounced enhancement of prominence for intermediate dephasing in the ensemble with low prominence.}
	\label{fig:prominence_split}
\end{figure}

Fig.~\ref{fig:prominence_split} shows two ensemble averages of this prominence as a function of the dephasing rate \(\gamma\),
extracted from an ensemble of $5000$ disorder realizations.
The blue (light) curve corresponds to the half of the realizations resulting in lower prominence for very weak dephasing with $\gamma=10^{-9}\,{\cal T}$,
whereas the black (dark) curve represents the ensemble average over the other half of realizations.
Both sub-ensembles share several features:
the overall prominence drops to zero for fast dephasing (\(\gamma \geq {\cal T}\)), confirming the absence of side-peaks in this limit.
The growth of the average prominence with decreasing dephasing rate up to $\gamma\simeq 10^{-2}\, {\cal T}$ is essentially the same in both sub-ensembles, and the prominence is independent of the dephasing rate $\gamma$ in the regime $\gamma< 10^{-4}\,{\cal T}$.
The central difference between both sub-ensembles lies in the maximum of prominence at around $\gamma \simeq 3 \cdot 10^{-3}\,{\cal T}$ in the ensemble with lower prominence, as opposed to the monotonic decrease in prominence with increasing dephasing rate $\gamma$ for the ensemble with higher prominence.

\section{Weak dephasing}

In the following, we will present a mostly analytic description for the emergence of side-peaks in the slow dephasing regime.
Based on this description we can then develop the physical understanding of the enhanced prominence for intermediate dephasing
shown in Fig.~\ref{fig:prominence_split} and the enhanced average population for intermediate dephasing at intermediate distances from the original peak shown in Fig.~\ref{fig:ensemble_avg}.

In order to obtain a weak dephasing approximation for the evolution of the system defined by the Lindblad master equation
\be
\dot{\rho} = -i[H,\rho] + \mathcal{L}[\rho],
\ee
with the Anderson Hamiltonian $H$ and the Lindbladian $\mathcal{L}$ taking the specific form
\be
\mathcal{L}[\rho] = -\gamma \sum_{x,y = 1}^{N} f(x,y) \ \prj{x}\rho\prj{y},
\ee
it is helpful to consider a transformation into the frame defined by the coherent dynamics $e^{-i H t}$.
The master equation for the transformed state $\tilde{\rho} = e^{i H t} \rho e^{-i H t}$
contains only terms proportional to $\gamma$ and reads
\be
\dot{\tilde{\rho}}=
\tilde{\cal L}_t[\tilde{\rho}]=
-\gamma \sum_{x,y = 1}^{N} f(x,y)P_x(t)\ \tilde\rho\ P_y(t),
\ee
with $P_z(t)=e^{i H t}\prj{z}e^{-i H t}$ for $z=x,y$.

For matrix elements in the eigenbasis $\ket{\Psi_i}$ of $H$ with eigenvalues $\varepsilon_i$, this implies
\be
\bra{\Psi_i}\dot{\tilde{\rho}}\ket{\Psi_j}= -\gamma \sum_{x,y,p,q = 1}^{N} f(x,y)\ \xi^t_{x y p q} \bra{\Psi_p}\tilde\rho\ket{\Psi_q},
\ee
with
\be
\xi^t_{x y p q} = e^{i(\varepsilon_i-\varepsilon_p+\varepsilon_q-\varepsilon_j)t} \ \Psi_i^\ast(x)\Psi_p(x)\Psi_q^\ast(y)\Psi_j(y),
\label{eq:expansion_coeff}
\ee
defined in terms of the short hand notation $\Psi(x)=\langle x|\Psi\rangle$.

In lowest order, {\it i.e.} linear in $\gamma$, the propagator $V(t)$ induced by $\tilde{\cal L}$ reads
\be
V(t)=\exp\left(\int_{0}^t d\tau\ \tilde{\cal L}_\tau+\mathcal{O}(\gamma^2)\right).
\ee
In the long-time limit, {\it i.e.} on time-scales that are long as compared to the relevant time-scales of the system Hamiltonian, this can be approximated by
\be
V(t)=\exp\left(\left(\lim_{T\to\infty}\frac{1}{T}\int_{0}^T d\tau\ \tilde{\cal L}_\tau\right)t+\mathcal{O}(\gamma^2)\right),
\ee
which justifies the definition of the effective Lindbladian
\be
\tilde{\cal L}_0=\lim_{T\to\infty}\frac{1}{T}\int_{0}^T d\tau\ \tilde{\cal L}_\tau.
\ee

Given the explicit form of the expansion coefficients in Eq.~\eqref{eq:expansion_coeff}, the integration results in a restriction of the terms for which the factor $\varepsilon_i-\varepsilon_p+\varepsilon_q-\varepsilon_j$ in the exponent vanishes.
Due to the random disorder in the Hamiltonian, one can neglect accidental cancellations.
For an off-diagonal element (with $i\neq j$) only the terms with $[p,q]=[i,j]$ contribute and one obtains
\be
\bra{\Psi_i}\tilde{\cal L}_0[\tilde{\rho}]\ket{\Psi_j}=-\gamma \ \mu_{ij} \ \bra{\Psi_i}\tilde\rho\ket{\Psi_j},
\ee
with
\be
\mu_{ij}  = \sum_{x,y = 1}^{N} f(x,y) \Psi_i^\ast(x)\Psi_i(x)\Psi_j^\ast(y)\Psi_j(y).
\ee
Off-diagonal elements in the eigenbasis of the system Hamiltonian thus decay exponentially.

For a diagonal element, the integration of the exponential function reduces the sum to the terms with $q=p$, and one therefore obtains the rate equation
\be
\bra{\Psi_i}\tilde{\cal L}_0[\tilde{\rho}]\ket{\Psi_i}=-\gamma \sum_{j = 1}^{N}\eta_{ij}\bra{\Psi_j}\tilde\rho\ket{\Psi_j},
\label{eq:slow_master}
\ee
with coupling elements,
\be
\eta_{ij}=
\sum_{x,y = 1}^{N} f(x,y) \Psi_i^\ast(x)\Psi_j(x)\Psi_j^\ast(y)\Psi_i(y),
\label{eq:coupling_parameter}
\ee
given in terms of overlaps between energy eigenstates $\ket{\psi_{i/j}}$ and site eigenstates $\ket{x/y}$, and the function $f(x,y)$ characterizing the ability of the Lindblad operator (Eq.~\eqref{eq:lindblad}) to resolve small spatial structures.
For sufficiently slow dephasing, an incoherent mixture of energy eigenstates will therefore remain such a mixture for all times, and the dephasing (in the site basis) only results in incoherent transitions between energy eigenstates.

The system thus decoheres in the eigenbasis of the system Hamiltonian, as it is typically the case in the weak dephasing limit, and the dephasing constant $\gamma$ enters only as multiplicative prefactor.
Its impact thus reduces to the definition of a time-scale, such that duration of growth and decay of interference structures depends on the actual value of $\gamma$, but quantities like maximal height of given structures (with the maximization performed over time) are indeed independent of $\gamma$, which explains why the prominence becomes independent of $\gamma$ for sufficiently weak dephasing in Fig.~\ref{fig:prominence_split}.

Transitions between states $\ket{\psi_{i}}$ and $\ket{\psi_{j}}$ can thus be sizeable if both states have substantial amplitudes on pairs of sites $\ket{x}$ and $\ket{y}$, and if the function $f(x,y)$ for those two sites is sufficiently large, which means that coherent superpositions of $\ket{x}$ and $\ket{y}$ decay quickly.

In the case of uniform dephasing ({\it i.e.} $f(x,y)=1$ for $x\neq y$), the latter condition is obsolete, such that generally many transitions between energy eigenstates are possible.
Any eigenstate will therefore decay into a mixture of many eigenstates, and fine structures that are contained in each eigenstate tend to average out in this mixture.
One would thus expect not to observe any pronounced side-peaks for this decoherence model, and we verified in numerical simulations that this is indeed the case.
If, on the other hand, the environment is limited in its resolution of small spatial structures, such that the function $f(x,y)$ is negligible
for small separations $|x-y|$, then only contributions with sufficiently large separation between $\ket{x}$ and $\ket{y}$ can contribute substantially to the sum in Eq.~\eqref{eq:coupling_parameter}.
Since, in addition, the transition amplitudes $\Psi_i^\ast(x) \Psi_i(y)$ become exponentially small with growing separation between $\ket{x}$ and $\ket{y}$\footnote{For weak disorder only states at the edge of the spectrum are exponentially localized, but since the system is initialized in its ground state, this statement holds true for all relevant transitions.}, only a very small number of terms contributes substantially to the sum in Eq.~\eqref{eq:coupling_parameter}.
The system state is thus typically given by a mixture of only few energy eigenstates,
and because of this restriction to a small number of eigenstates, spiked structures do not necessarily average out, but side-peaks can arise.
The dependence on the transition amplitude $\braket{x | \psi_i}\braket{\psi_i | y}$ asserts that side-peaks will arise predominantly in the vicinity of the tails of the exponentially localized initial state, which is exactly what can be observed in Fig.~\ref{fig:example_evolution}.

In order to draw further conclusions from Eq.~\eqref{eq:coupling_parameter} that are independent of a specific realization of disorder,
one can approximate eigenstates of the Anderson Hamiltonian as plane waves, with the same wave number as obtained for the Hamiltonian without disorder~\cite{doi:10.1063/1.4797477}, modulated by an envelope that imprints the exponential localization.
Such an ansatz permits to analytically evaluate the coupling constants $\eta_{ij}$ given in Eq.~\eqref{eq:coupling_parameter} in the continuum limit, where the summation is replaced by an integral.

The states $\ket{\chi_j}$ with the associated wave functions
\be
\chi_{j}(x)=\ovl{x}{\chi_j}=\sqrt{\lole_j}e^{-\lole_j |x-x_j|}e^{ik_jx},
\ee
exponentially localized with inverse localization length $\lole_j$ around center $z_j$, and with wave number $k_j$ capture the essential features of eigenstates of the Anderson Hamiltonian.
The only -- but crucial -- feature that is missing, is that two states $\ket{\chi_i}$ and $\ket{\chi_j}$ are not necessarily orthogonal to each other.

A pair of orthogonal states with well-defined localization properties is given by $\ket{\chi}=\ket{\chi_1}$ and
\be
\ket{\chi_\perp}=\frac{1}{\sqrt{N}}
\left(\frac{\ket{\chi_{2}}}{\ovl{\chi_{1}}{\chi_{2}}}-\frac{\ket{\chi_{3}}}{\ovl{\chi_{1}}{\chi_{3}}}\right),
\ee
with $\lole_3=\lole_2$ and $z_3=z_2$,
and the normalization constant
\begin{multline}
	N = \frac{1}{|\ovl{\chi_{1}}{\chi_{2}}|^2}+ \frac{1}{|\ovl{\chi_{1}}{\chi_{3}}|^2} \\
	  - \frac{\ovl{\chi_{2}}{\chi_{3}}}{\ovl{\chi_{2}}{\chi_{1}}\ovl{\chi_{1}}{\chi_{3}}}-
\frac{\ovl{\chi_{3}}{\chi_{2}}}{\ovl{\chi_{3}}{\chi_{1}}\ovl{\chi_{1}}{\chi_{2}}}.
\end{multline}
We are interested in the function
\be
F_q=\int dx\ dy\ f_q(x-y)\ \chi(x)\chi^\ast(y)\chi_\perp(y)\chi_\perp^\ast(x),
\ee
with $f_0=1-\delta_{x-y}$ and $f_1=|x-y|$, that determines the coupling between the states $\ket{\chi}$ and $\ket{\chi_\perp}$.

Evaluation of $F_q$ requires evaluation of elementary integrals of the form $\int dx\ x^q\exp(ikx)$ only, so that $F_q$ can readily be expressed as an analytic function of wave numbers, localization lengths and centers.
Physical insight, however, is not necessarily obtained from the general dependencies, but from limiting cases.
Expressing wave vector differences as $k_2-k_1=k-\frac{dk}{2}$ and $k_3-k_1=k+\frac{dk}{2}$ in terms of central have number $k$ and relative wave number $dk$, permits to take the limit
\be
F_q^{(0)}=\lim_{dk\to 0}F_q.
\ee
Assuming the states $\ket{\chi}$ and $\ket{\chi_\perp}$ to have the localization lengths $\lole_1=\lole_2/2=\lole$, the limits $F_q^{(0)}$
are functions of the two independent variables $r=\lole |z_1-z_2|$ and $\kappa=k/\lole$ only, and help to understand the features identified in the explicit solutions of the Anderson Hamiltonian.

\begin{figure*}
	\centering
	\input{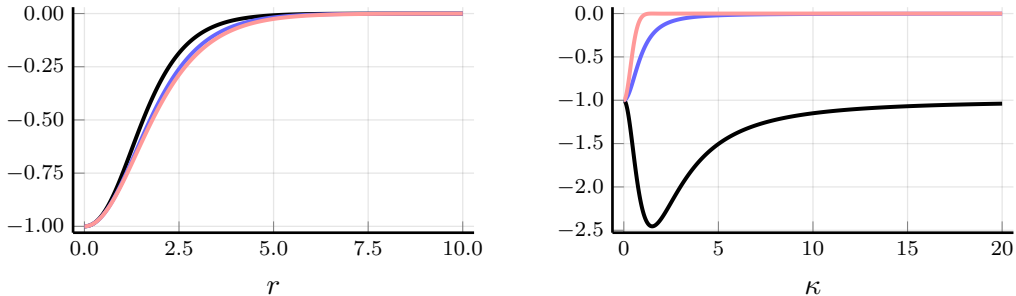}
	\caption{The left plot depicts the coupling constant $F_q^{(0)}(r,\kappa=0)/|F_q^{(0)}(r=0,\kappa=0)|$ between two states $\ket{\chi}$ and $\ket{\chi_\perp}$ as function of their separation of localization center in the limit of vanishing difference in wave number. The inverse localization length for $\ket{\chi_\perp}$ is chosen as $\lole_2=\lole_1/2$. Black (dark) denotes $F_0$, blue (medium) denotes $F_1$ and red (light) denotes $F_2$.
The right plot depicts the same coupling constants as function of $\kappa$ with $r=0$ and the same other parameters as the left plot.}
\label{fig:analytic_couplings_dependece}
\end{figure*}

Fig.~\ref{fig:analytic_couplings_dependece} depicts $F_q^{(0)}(r=0,\kappa)/|F_q^{(0)}(r=0,\kappa=0)|$ (left) and $F_q^{(0)}(r,\kappa=0)/|F_q^{(0)}(r=0,\kappa=0)|$ (right) for $q=0$ (depicted in black (dark)), $q=1$ (blue (medium)) and $q=2$ (red (light)).
Since there is no meaningful comparison of the ordinates for the different cases of $q$, only normalized, and thus unitless versions of the functions $F_q^{(0)}$ are shown.
Asymptotically, the decay with $r$ is dominated by the exponential decay $\exp(-2r)$, but as one can see on the left of Fig.~\ref{fig:analytic_couplings_dependece}, the dependence of $F_q^{(0)}$ on $r$ is largely independent of $q$, {\it i.e.} the decoherence model, for any value of $r$.
This is fundamentally different for the dependence on $\kappa$ depicted on the right of Fig.~\ref{fig:analytic_couplings_dependece}.
Here, one can see that uniform decoherence ({\it i.e.} $q=0$) results in
substantial coupling for any value of $k$, whereas the coupling decays rapidly with increasing $k$ for $q>0$.
The coupling between eigenstates thus becomes increasingly restricted to spectrally close eigenstates with growing $q$, which explains why side-peaks become particularly pronounced with increasing $q$.

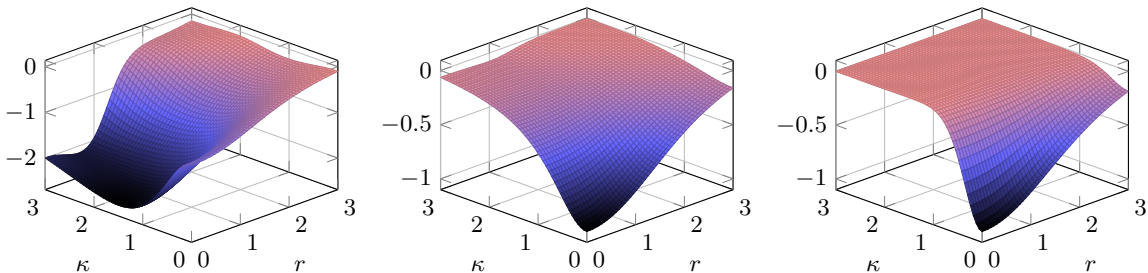
\begin{figure*}
	\centering
	\def\scale{0.34}
\begin{tikzpicture}
  \begin{axis}[view={-45}{30}, grid=both, xlabel = {$r$}, ylabel = {$\kappa$},
               xtick = {0, 1, 2, 3}, xticklabels = {$0$,$1$,$2$,$3$},
               ytick = {0, 1, 2, 3}, yticklabels = {$0$,$1$,$2$,$3$},
               colormap={colmap}{rgb=(\colorared,\coloragreen,\colorablue) rgb=(\colorbred,\colorbgreen,\colorbblue) rgb=(\colorcred,\colorcgreen,\colorcblue)},
               width = {\scale*\textwidth}]
    \addplot3[surf] file {F_q_0_normalised.txt};
  \end{axis}
\end{tikzpicture} \begin{tikzpicture}
  \begin{axis}[view={-45}{30}, grid=both, xlabel = {$r$}, ylabel = {$\kappa$},
               xtick = {0, 1, 2, 3}, xticklabels = {$0$,$1$,$2$,$3$},
               ytick = {0, 1, 2, 3}, yticklabels = {$0$,$1$,$2$,$3$},
               colormap={colmap}{rgb=(\colorared,\coloragreen,\colorablue) rgb=(\colorbred,\colorbgreen,\colorbblue) rgb=(\colorcred,\colorcgreen,\colorcblue)},
               width = {\scale*\textwidth}]
    \addplot3[surf] file {F_q_1_normalised.txt};
  \end{axis}
\end{tikzpicture} \begin{tikzpicture}
  \begin{axis}[view={-45}{30}, grid=both, xlabel = {$r$}, ylabel = {$\kappa$},
               xtick = {0, 1, 2, 3}, xticklabels = {$0$,$1$,$2$,$3$},
               ytick = {0, 1, 2, 3}, yticklabels = {$0$,$1$,$2$,$3$},
               colormap={colmap}{rgb=(\colorared,\coloragreen,\colorablue) rgb=(\colorbred,\colorbgreen,\colorbblue) rgb=(\colorcred,\colorcgreen,\colorcblue)},
               width = {\scale*\textwidth}]
    \addplot3[surf] file {F_q_2_normalised.txt};
  \end{axis}
\end{tikzpicture}
	\caption{$F_q^{(0)}(r,\kappa)/|F_q^{(0)}(r=0,\kappa=0)|$ for $q=0$ (left), $q=1$ (center) and $q=2$ (right) with the same parameter values as in Fig.~\ref{fig:analytic_couplings_dependece}.}
	\label{fig:analytic_couplings_surface}
\end{figure*}

The behavior observed in Fig.~\ref{fig:analytic_couplings_dependece} is confirmed in Fig.~\ref{fig:analytic_couplings_surface} where $F_q^{(0)}(r,\kappa)/|F_q^{(0)}(r=0,\kappa=0)|$ is depicted as function of both $r$ and $\kappa$ for $q=0$ (left), $q=1$ (center) and $q=2$ (right).
As one can see, the case of uniform dephasing $q=0$ results in substantial coupling independently of $\kappa$ for sufficiently small distances $r$ between two states, whereas the cases $q=1$ and $q=2$ results in enhanced coupling between spectrally close states ({\it i.e.} small $\kappa$) also for finite, small values of $r$.

The observations made in Figs.~\ref{fig:analytic_couplings_dependece} and~\ref{fig:analytic_couplings_surface} can also be confirmed in limiting cases for which the functions $F_q^{(0)}$ have a reasonably simple dependence on wave-numbers and localization centers.
In the case of short distances $z=z_1-z_2$ between the localization centers and low differences in the wave numbers,
the functions $F_q^{(0)}$ have the dependence $F_q^{(0)}\propto k^{-2q}$ for $q=0,1,$ and $2$.

In the case of uniform dephasing, there is thus a finite coupling $F_0^{(0)}$, even for states with large difference in wave number, whereas the coupling between states with different wave-numbers is suppressed for the other dephasing models.

\begin{figure}[h]
	\centering
\input{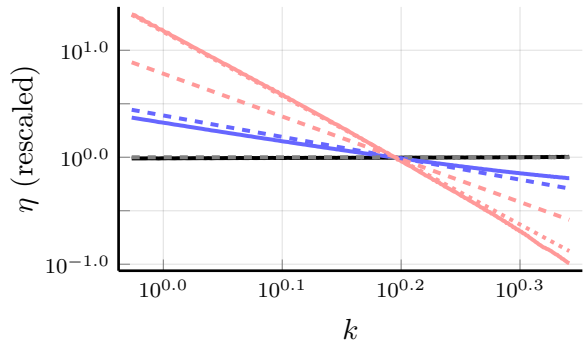}
\caption{Normalised ensemble average of the coupling constants (Eq.~\eqref{eq:coupling_parameter}) over 50000 disorder realizations as a function of the wave number distance from the ground state for states in the center of the energy spectrum, with the normalization chosen such that the averages coincide for the central wave number difference at $\eta=1$.
	The black (dark) curve corresponds to the uniform dephasing model, blue (medium) represents the linear model with $q=1$ and red (light) the quadratic model ($q=2$).
	The dashed grey (dark) line represents a constant relation between averaged coupling constants and $k$, the dashed blue (medium) line shows a decay of the coupling constants according to $1/k^2$, the dashed red (light) line a decay according to $1/k^4$ and the dotted red (light) line a decay according to $1/k^6$.}
	\label{fig:couplings_power_law}
\end{figure}

These power law suppressions for large $k$ also manifest themselves in the ensemble average of the actual coupling constants between the ground state and higher excited states over multiple disorder realizations as shown in Fig.~\ref{fig:couplings_power_law}.
The coupling constants are depicted for states around the center of the spectrum for which the localization length is approximately constant so that the dependence of the coupling constants on this can be neglected. The respective wave number difference for the $i$-th excited energy eigenstate is taken to be the value obtained for the Hamiltonian without disorder, $k_i = i \pi/(N+1)$~\cite{doi:10.1063/1.4797477}.
As can be seen, the averaged coupling constants are approximately constant across the depicted region for the uniform dephasing model and decay for anisotropic dephasing.
For the linear model, the power law decay according to $\sim 1/k^2$ as obtained for the continuum limit describes the dominating contribution for the relationship between coupling constant and difference in wave number.
For the quadratic model however, a higher order power law decay according to $\sim 1/k^6$ appears to dominate over the approximated decay $\sim 1/k^4$.
This can be attributed to the fact that the $\sim 1/k^6$-component is not sufficiently decayed in the range depicted in Fig.~\ref{fig:couplings_power_law} so that the leading order term $\sim 1/k^4$ can not be clearly identified.
In physical terms, this implies
an even stronger selection of energy eigenstates which couple significantly to the ground state.

Following from the analytic estimates and the numerical observations, we obtain a clear relationship between the emergence of side peaks in the weak dephasing limit and the eigenstates of $H$.
Consistently with observations in the disordered systems of finite size, the increasing suppression of couplings between states with large difference in wave number $k$ implies that transitions are expected to be sizeable only between spectrally close states.

Together with the above findings, this defines the rather stringent condition for the emergence of side peaks:

{\it a prominent side-peak can only occur if there is a strongly peaked eigenstate that is spectrally close to the initial state, and both of these states have large transition amplitudes $\braket{x | \circ }\braket{\circ | y}$ for pairs of sites $x$ and $y$ with sizeable decoherence rate.}

\section{Intermediate dephasing}

If the dephasing constant is too large for the lowest order approximation to hold, the basis in which the system decoheres is typically not the Hamiltonian eigenbasis anymore.
The preferred basis rather depends on the value of $\gamma$, and in the limiting case of strong decoherence it tends towards the site-basis $\ket{x}$, {\it i.e.} the basis that is selected by the underlying Lindbladian ${\cal L}$.

Intermediate dephasing as shown in Fig.~\ref{fig:example_evolution}(c) is realized for decoherence rates that are too large to admit the treatment of Eq.~\eqref{eq:slow_master} but still small enough for some coherent dynamics to occur, and in the strong dephasing regime the system dynamics is essentially incoherent.
For sufficiently large dephasing rates (roughly $\gamma \gtrsim 10^{-6} \, \mathcal{T}$), the system state can no longer be described as a purely incoherent mixture of energy eigenstates, but coherences between energy eigenstates can build up.
The first observable consequence of this is that the overall dynamics slows down~\cite{PhysRevB.92.014203}, such that it takes longer for the initial peak  to dissolve into the flat steady state distribution.
This resembles the onset of a quantum Zeno effect~\cite{doi:10.1063/1.523304}, even though decoherence is still too weak for complete suppression of population dynamics.
Populations can therefore still propagate through the chain, but, as shown in Fig.~\ref{fig:ensemble_avg}, the propagation over large distances is suppressed as compared to the weak dephasing regime.
Since dynamics on shorter scales has a more coherent character than dynamics on larger scales, populations can thus propagate so that an initially localized peak can start to dissolve, resulting in the fine interference structures that are apparent in Fig.~\ref{fig:example_evolution}.
The onset of quantum Zeno dynamics then tends to prevent such a structure to expand over a larger spatial region.
This suppression, in turn, implies an enhanced probability to remain in the vicinity of the initial distribution, which is reflected by the increased probability $\bar P$ for sites $270\lesssim x\lesssim 290$ for intermediate dephasing in Fig.~\ref{fig:ensemble_avg}.

This onset of quantum Zeno dynamics also explains the maximum of prominence for intermediate dephasing rates in Fig.~\ref{fig:prominence_split}.
In the sub-ensemble with less prominence, there tend to be several side peaks with similar prominence in the weak dephasing regime,
and the growth of several peaks imposes limits on the growth of each individual peak.
For intermediate dephasing, however, the growth of peaks further away from the initial peak is suppressed by the onset of quantum Zeno dynamics.
Since dephasing is not strong enough to suppress the growth of all side-peaks, this results in a more focused flow of population to selected peaks.

\section{Conlusions \& outlook}

The rise and decay of interference structures indicates neatly the interplay of interference and decoherence.
For the sake of specificity most of the explicit situations shown here correspond to the function $f(x,y)=|x-y|$, but none of the results discussed here are specific to this particular model.
We found qualitatively the same results for $f_q(|x-y|)=|x-y|^q$ with $q=2$ and $3$, with the only difference that side-peaks get more pronounced with increasing $q$, which is consistent with the observation that the spectral restriction becomes more stringent with increasing $q$.

The unavoidable overhead required for the simulation of open quantum systems makes generalizations to two- or three-dimensional systems~\cite{PhysRevLett.102.106406} prohibitively expensive for simulations on classical computers.
Quantum simulators~\cite{RevModPhys.86.153} on the other hand can be viable options.
Analogue optical quantum simulators were already used to simulate transport in disordered landscapes~\cite{Harris:2017fk}.
Atomic platforms with optical lattices~\cite{Nature_453_891,roati2008anderson} realize predominantly coherent dynamics, but temporally modulated site energies can introduce dephasing with a quadratically growing dephasing rate ({\it i.e.} $\sim|x-y|^2$)~\cite{huntergordon2019quantum}.
Digital quantum simulations with currently existing hardware~\cite{Linke3305} comprised of several tens of qubits suffer inherently from decoherence, and the quantum simulation of the dephasing Anderson model would be perfectly suited to turn such a limitation into an asset.

{\it Acknowledgement} This work benefited greatly from stimulating discussions with Yoshitaka Tanimura, Rob Nyman and Johannes Knolle.
Financial support by JSPS in terms of the fellowship S16025 and hospitality by Kyoto University are gratefully acknowledged.

\bibliographystyle{apsrev4-1}
\bibliography{manuscript_rath}
\end{document}